\documentstyle[aps,tighten,epsfig]{revtex} 
\begin{document} \draft

\title{ TIME-DEPENDENT GINZBURG-LANDAU MODEL WITH NOISE \\
                            FOR                         \\
                NEUTRAL S-WAVE SUPERCONDUCTOR      }

\author{ Jacek Dziarmaga
         \thanks{dziarmaga@t6-serv.lanl.gov} }
\address{ Los Alamos National Laboratory,
          Theoretical Astrophysics T-6,
          MS-B288, Los Alamos, NM 87545, USA \\
                         and \\
          Institute of Physics, Jagiellonian University,
          Krak\'ow, Poland  }
\date{ May 4, 1999 }

\maketitle
\tighten

\begin{abstract} \noindent 
Time-dependent Ginzburg-Landau model with noise for neutral s-wave
superconductor is derived from real-time finite-temperature quantum field
theory by integrating over fermions in the density matrix propagator.
Quantum decoherence due to environment of pair breaking fluctuations is
identified. Dynamics is described by Langevin equation in the near
critical regime and by a nonlinear Schrodinger equation at zero
temperature.
\end{abstract}

\pacs{PACS 74.20.Fg 74.40.+k 74.20.De }

\section*{ BACKGROUND }

  Ever since the advent of the Ginzburg-Landau theory of superconductivity
\cite{gl} there was continuous effort to derive it from the microscopic
theory. The first step was made by Gorkov \cite{gorkov} in the static
case. The static case is of much thermodynamical interest, see
e.g.\cite{stone} for recent results. However, dynamical out of equilibrium
phenomena require a time-dependent version of the Ginzburg-Landau model.
This include for example formation of topological defects during a fast
second order phase transition \cite{zurek}. Models of defect formation
normally assume the dynamics is described in the near-critical regime by a
classical Langevin-type partial differential equation with external white
noise. First derivation of a time-dependent Ginzburg-Landau model was
given in \cite{at}, similar approach is described in the recent review
\cite{schakel}. Neither of these works explicitly discusses noise or
decoherence of the order parameter field. This paper aims at filling this
gap for the simplest example of neutral s-wave superconductor.

  A topic in its own right is an effective quantum action at zero
temperature. Natural expectation is a kind of nonlinear Schrodinger
equation analogous to the Gross-Pitaevskii \cite{gp} equation for
Bose-Einstein condensates. Such equations were indeed constructed
\cite{ao,stone2} for the phase field or in perturbative regime. An
effective {\it nonperturbative} Lagrangian for the pair field $\Delta$ is
one of our results.

  We construct the density matrix propagator for electrons with the help
of the finite temperature Kyeldysh contour. The Hubbard-Stratonovich
transformation is performed to introduce the order parameter $\Delta$ and
at the same time to make electronic Hamiltonian quadratic. Next, we
integrate over fermions and obtain effective description in terms of the
auxillary $\Delta$ field. From the effective action in terms of $\Delta$'s
living on the time-ordered and anti-time-ordered branch of the contour one
can read the effective equation satisfied by the order parameter including
dissipation and noise. We identify decoherence due to interaction of the
pair field with pair-breaking fermionic fluctuations.

  Our paper is organized as follows. In Section {\bf I} we define the
model and perform Hubbard-Stratonovich transformation to introduce the
pair field $\Delta$. Our aim is to integrate out fermions in the density
matrix propagator and read from it an effective action for $\Delta$. High
temperature case is considered in Section {\bf II}. In this regime the
expectation value of the order parameter is zero. In the spirit of the
Ginzburg-Landau model we make an expansion in powers of $\Delta$ to
obtain, in the long wavelength limit, partial differential Langevin
equation with white noise. At the same time the long wavelength modes of
$\Delta$ are subject to quantum decoherence and thus they can be
considered as classical. This equation can be extended below the critical
temperature $T_c$ to describe spinodal decomposition as long as
temperature remains greater than the developing gap. Further below $T_c$
the gap in the quasiparticle spectrum can no longer be neglected. In
Section {\bf III} we perform an expansion around uniform equilibrium order
parameter. The fluctuations decohere, their evolution is effectively
classical. They satisfy integro-differential Langevin equation with
colored noise and nonlocal dissipation kernel. Properties of the system
change qualitatively as we approach zero temperature, see Section {\bf
IV}. Some sources of noise and decoherence are suppressed when temperature
becomes much lower than the gap, other are restricted only to frequencies
greater than twice the gap. Thus in the low temperature and long
wavelength limit there is no noise or decoherence, we obtain an effective
quantum Lagrangian, which leads to a kind of nonlinear Schrodinger
equation for the order parameter $\Delta$.

\section{ THE MODEL }

  S-wave neutral superconductor is described by the Lagrangian 

\begin{equation}\label{Lpsi}
L_e[\psi] = 
\int d^3x\; \left[
\sum_{\alpha=\uparrow,\downarrow}
\left(
i \psi^{\dagger}_{\alpha} \partial_t \psi_{\alpha} +
\frac{1}{2m} \psi^{\dagger}_{\alpha} \nabla^2 \psi_{\alpha} +
\mu \psi^{\dagger}_{\alpha} \psi_{\alpha} 
\right) +
V  
\psi^{\dagger}_{\uparrow} \psi_{\uparrow}
\psi^{\dagger}_{\downarrow} \psi_{\downarrow}
\right]                                            \;.
\end{equation}
$\psi_{\uparrow / \downarrow}$ are spin up/down electronic Grassman
fields. $V$ is a coupling constant for the s-wave pairing between spin-up
and spin-down electrons. By the standard Hubbard-Stratonovich
transformation the Lagrangian is {\it equivalent} to

\begin{equation}\label{L}
L[\psi,\Delta]=
\int d^3x\; \left[
\sum_{\alpha=\uparrow,\downarrow}
\left(
i \psi^{\dagger}_{\alpha} \partial_t \psi_{\alpha} +
\frac{1}{2m} \psi^{\dagger}_{\alpha} \nabla^2 \psi_{\alpha} +
\mu \psi^{\dagger}_{\alpha} \psi_{\alpha}
\right) -
\frac{(2\pi)^3}{V} \Delta^{\star}\Delta +
(2\pi)^{3/2}\Delta^{\star}\psi_{\downarrow}\psi_{\uparrow} +
(2\pi)^{3/2}\Delta\psi^{\dagger}_{\uparrow}\psi^{\dagger}_{\downarrow}
\right]                                            \;.
\end{equation}
Functional integration over the auxillary $\Delta$ leads back to $L_e$.
Variation of $L$ with respect to the pair field $\Delta^{\star}$ gives

\begin{equation}\label{pair}
\Delta=
\frac{V}{(2\pi)^{3/2}} \; \psi_{\downarrow}\psi_{\uparrow} \;.
\end{equation}
{\bf Note that our $\Delta$ differs from the standard definition by the
factor $(2\pi)^{-3/2}$ }. Our goal is to integrate over
$\psi^{\dagger},\psi$ to obtain effective theory in terms of pair fields.
We split this task into subcases.

\section{ ABOVE $T_c$ }

 The equilibrium value of $\Delta$ vanishes above $T_c$, we perform
expansion around equilibrium in powers of $\Delta$. It is convenient to
introduce Fourier transforms in space

\begin{eqnarray}\label{Fourier}
\psi(t,\vec {x})&=&
\int d^3k\; \frac{e^{i\vec{k}\vec{x}}}{(2\pi)^{3/2}} \;
            a(t,\vec{k})                           \;\;,\nonumber\\
\Delta(t,\vec{x})&=&
\int d^3k\; \frac{e^{i\vec{k}\vec{x}}}{(2\pi)^{3/2}} \;
            \Delta(t,\vec{k})                      \;\;.
\end{eqnarray}
For clarity we denote the transform of $\Delta$ by the same letter. In 
Fourier space the Lagrangian reads

\begin{eqnarray}\label{LFourier}
&& L[a,\Delta]=L_0[a]+L_{\Delta}[\Delta]+L_{int}[a,\Delta] \;\;,\\
&& L_0[a]=
   \int d^3k\; \left[
   \sum_{\alpha=\uparrow,\downarrow}
   \left(  
   i a^{\dagger}_{\alpha} \partial_t a_{\alpha} -
   \epsilon_{\vec{k}} a^{\dagger}_{\alpha} a_{\alpha} 
   \right) \right]                      \;\;,\nonumber\\ 
&& L_{\Delta}[\Delta]=
   \int d^3k\; \left[
   -\frac{\Delta^{\star}\Delta}{V} \right] \;\;,\nonumber\\
&& L_{int}[a,\Delta]=
   \int d^3k\; \left[
    \Delta^{\star} F_{\vec{k}}[a]  +
    \Delta         F^{\dagger}_{\vec{k}}[a]
    \right]                             \;\;,\nonumber
\end{eqnarray}
where $\epsilon_{\vec{k}}=\frac{k^2}{2m}-\mu$ and the interaction
vertex is

\begin{equation}\label{F}
F_{\vec{k}}[a]= \int d^3p\;
a_{\downarrow}(t,\vec{p}) a_{\uparrow}(t,\vec{k}-\vec{p}) \;\;.
\end{equation}

The density matrix propagator for $\Delta$ can be obtained by integrating
over fermions. To this end we put the fields on the Kyeldysh contour. In
the complex time plane the contour runs from $t_i$ to $t_f$ (time ordered
branch), then back from $t_f$ to $t_i$ (anti-time ordered branch) and from
$t_i$ to $t_i-i\beta$, where $\beta=1/k_B T$ (imaginary branch). The
imaginary branch prepares the system at $t_i$ in the state of thermal and
chemical equilibrium characterized by $\beta$ and $\mu$. When $t_{f/i}$
is moved to $\pm\infty$, we are left with fields living on the
time-ordered branch ($a_+,\Delta_+$) and those on the anti-time-ordered
branch ($a_-,\Delta_-$). In terms of these the density matrix propagator
can be written as

\begin{eqnarray}\label{J}
&&\int D\Delta^{\star}_+ D\Delta_+ D\Delta^{\star}_- D\Delta_-
  e^{i\int dt\; \left( L_{\Delta}[\Delta_+]
                     - L_{\Delta}[\Delta_-] \right)}
  e^{ iS_F[\Delta_+,\Delta_-] }           \;\;,\nonumber\\
&& e^{ iS_F[\Delta_+,\Delta_-] }=
   \int Da^{\dagger}_+ Da_+ Da^{\dagger}_- Da_-
   e^{i\int dt\; \left( L_0[a_+]+L_{int}[a_+,\Delta_+] 
                      - L_0[a_-]-L_{int}[a_-,\Delta_-] \right)} \;\;,
\end{eqnarray}
In the absence of $\Delta$ the free fermionic operators of {\it the same
spin} have propagators

\begin{eqnarray}\label{<>}
&&\langle \hat T [\; a_+(t,\vec{k}) \; 
                     a_+^{\dagger}(t',\vec{k}') \;]\rangle_{\vec{v}}=
  \delta(\vec{k}-\vec{k}')
  e^{-i[ \epsilon_{\vec{k}} + \vec k \vec v + \frac{mv^2}{2} ](t-t')}
  \left[ \theta(t-t')f(-\beta\epsilon_{ \vec{k}-m\vec{v} }) 
        -\theta(t'-t)f(\beta\epsilon_{ \vec{k}-m\vec{v} } ) \right]
                                                            \;,\nonumber\\
&&\langle \hat T [\; a_-(t,\vec{k}) \; 
                     a_-^{\dagger}(t',\vec{k}') \;]\rangle_{\vec{v}}=  
  \delta(\vec{k}-\vec{k}')
  e^{-i[ \epsilon_{\vec{k}} + \vec k \vec v + \frac{mv^2}{2} ](t-t')}
  \left[ \theta(t'-t)f(-\beta\epsilon_{ \vec{k}-m\vec{v} })  
        -\theta(t-t')f(\beta\epsilon_{ \vec{k}-m\vec{v} }) \right]
                                                           \;,\nonumber\\
&&\langle \hat T [\; a_-(t,\vec{k}) \; 
                     a_+^{\dagger}(t',\vec{k}') \;]\rangle_{\vec{v}}=
  \delta(\vec{k}-\vec{k}')
  e^{-i[ \epsilon_{\vec{k}} + \vec k \vec v + \frac{mv^2}{2} ](t-t')}
  f(-\beta\epsilon_{\vec{k}-m\vec{v}}) \;,\nonumber\\
&&\langle \hat T [\; a_+(t,\vec{k}) \; 
                     a_-^{\dagger}(t',\vec{k}') \;]\rangle_{\vec{v}}=
  - \delta(\vec{k}-\vec{k}')
  e^{-i[ \epsilon_{\vec{k}} + \vec k \vec v + \frac{mv^2}{2} ](t-t')}
  f(\beta\epsilon_{\vec{k}-m\vec{v}}) \;.
\end{eqnarray}
$f(x)=(1+e^x)^{-1}$ comes from the Fermi-Dirac distribution and $\hat T$
means ordering along the Kyeldysh contour. $\langle\dots\rangle_{\vec{v}}$
means an average over the initial state which is a state of thermal
equilibrium in a reference frame moving with velocity $\vec{v}$. We take
such a generalized initial distribution to keep trace of Galilean
symmetry.

The influence action $S_F$ in the influence functional $\exp{iS_F}$,
compare Eq.(\ref{J}), can not be found exactly because it is quadratic in
the fermionic fields. Instead we work out $S_F$ by perturbative expansion
in powers of the order parameter $\Delta,\Delta^{\star}$. This expansion
is in the spirit of the Ginzburg-Landau model. As $ \langle F_{\vec{k}}[a]
\rangle_{\vec{v}} = 0 $ the leading order term is quadratic

\begin{eqnarray}\label{S2}
S_{F}^{(2)}[\Delta_+,\Delta_-]=
       \frac{i}{2}\int dtdt'\;\int d^3kd^3k'\;  
       (\;[\; 
          &&2\Delta_+(t,\vec{k}) \; 
             \langle\hat T[\; F^{\dagger}_{\vec{k}}[a_+(t)] \;
                            F_{\vec{k}'}[a_+(t')]  \;]\rangle_{\vec{v}}\;
             \Delta_+^{\star}(t',\vec{k}')                 \nonumber\\
          &&-\Delta_+(t,\vec{k}) \;
             \langle\hat T[\; F^{\dagger}_{\vec{k}}[a_+(t)] \;
                            F_{\vec{k}'}[a_-(t')]  \;]\rangle_{\vec{v}}\;
             \Delta_-^{\star}(t',\vec{k}')                 \nonumber\\
          &&-\Delta_+^{\star}(t,\vec{k}) \;
             \langle\hat T[\; F_{\vec{k}}[a_+(t)] \;
                            F^{\dagger}_{\vec{k}'}[a_-(t')]\;]
             \rangle_{\vec{v}} \;
             \Delta_-(t',\vec{k}')    \;]                  \nonumber\\
          && +\;\;\; [ + \leftrightarrow - ] \;\;\;)   \;\;.
\end{eqnarray}
This expression can be worked out with the help of the correlators
(\ref{<>}) to give

\begin{eqnarray}\label{}
&&S_{F}^{(2)}[\Delta_+,\Delta_-]=
  \sum_{a,b=+,-} 
  \int d\omega d^3k\;
  \Delta_a(\omega,\vec{k}) \; \Delta_b^{\star}(\omega,\vec{k}) \;
  G_{ab}(\omega,\vec{k})  \;,\\
&& G_{++}(\omega - 2\vec k \vec v -mv^2,\vec{k}-2m\vec{v})=
   \int d^3p 
   \left[
   \frac{ f(-\beta\epsilon_{\vec{p}}) f(-\beta\epsilon_{\vec{k}-\vec{p}})}
        { \omega+\epsilon_{\vec{p}}+\epsilon_{\vec{k}-\vec{p}}-i\eta }
  -\frac{ f(\beta\epsilon_{\vec{p}}) f(\beta\epsilon_{\vec{k}-\vec{p}})}
        { \omega+\epsilon_{\vec{p}}+\epsilon_{\vec{k}-\vec{p}}+i\eta }
   \right]       \;,\nonumber\\
&& G_{--}(\omega - 2\vec k \vec v -mv^2,\vec{k}-2m\vec{v})=
   \int d^3p
   \left[
   \frac{ f(\beta\epsilon_{\vec{p}}) f(\beta\epsilon_{\vec{k}-\vec{p}})}
        { \omega+\epsilon_{\vec{p}}+\epsilon_{\vec{k}-\vec{p}}-i\eta }  
  -\frac{ f(-\beta\epsilon_{\vec{p}}) f(-\beta\epsilon_{\vec{k}-\vec{p}})}
        { \omega+\epsilon_{\vec{p}}+\epsilon_{\vec{k}-\vec{p}}+i\eta }
   \right]       \;,\nonumber\\
&& G_{+-}(\omega - 2\vec k \vec v -mv^2,\vec{k}-2m\vec{v})=
   -2\pi i \int d^3p\;
   \delta(\omega+\epsilon_{\vec{p}}+\epsilon_{\vec{k}-\vec{p}})\;
   f(-\beta\epsilon_{\vec{p}}) 
   f(-\beta\epsilon_{\vec{k}-\vec{p}}) \;\\
&& G_{-+}(\omega - 2\vec k \vec v -mv^2,\vec{k}-2m\vec{v})=
   -2\pi i \int d^3p\;
   \delta(\omega+\epsilon_{\vec{p}}+\epsilon_{\vec{k}-\vec{p}})\;
   f(\beta\epsilon_{\vec{p}})
   f(\beta\epsilon_{\vec{k}-\vec{p}}) \;
\end{eqnarray}
Where $\eta$ is an infinitesimal positive parameter.
$G_{ab}(\omega,\vec{k})$ depends on $\vec{v}$ through $\omega+ 2\vec k
\vec v + \frac{mv^2}{2}$ and $\vec{k}+2m\vec{v}$ which means that $\Delta$
transforms under Galilean symmetry like mass $2m$ pair field. We introduce
an average and relative field, which are the usual coordinates used in
Wigner transform,

\begin{equation}\label{deltaR}
\Delta_{\pm}\;=\;\Delta\;\pm\;\frac{R}{2} \;.
\end{equation}
With this change of variables and with the identity
$(y+i\eta)^{-1}=P.V.\;y^{-1}-i\pi\delta(y)$ we obtain the quadratic part
of the effective action, which contains both real and imaginary parts

\begin{eqnarray}\label{S2eff}
&&S_{eff}^{(2)}[\Delta,R]=
  S_{\Delta}[\Delta,R]+S_F^{(2)}=    \nonumber\\
&&= \int d\omega d^3k\;
    \left\{ (\Delta R^{\star}+\Delta^{\star}R)
            [-\frac{(2\pi)^3}{V} + G(\omega,\vec{k})] \;+\;
            i(\Delta^{\star}R-\Delta R^{\star})\;D(\omega,\vec{k}) \;+\;
            iR^{\star}R\;K(\omega,\vec{k}) 
\right\}  \\
&& G(\omega - 2\vec k \vec v -mv^2,\vec{k}-2m\vec{v})=
   P.V.\int d^3p\;
\frac{ f(-\beta\epsilon_{\vec{p}}) f(-\beta\epsilon_{\vec{k}-\vec{p}}) -
       f(\beta\epsilon_{\vec{p}}) f(\beta\epsilon_{\vec{k}-\vec{p}})     }
     { \omega+\epsilon_{\vec{p}}+\epsilon_{\vec{k}-\vec{p}}              }
                                                            \nonumber\\
&& K(\omega - 2\vec k \vec v -mv^2,\vec{k}-2m\vec{v})=\pi\int d^3p\;
   \delta(\omega+\epsilon_{\vec{p}}+\epsilon_{\vec{k}-\vec{p}})\;
   [f(\beta\epsilon_{\vec{p}}) f(\beta\epsilon_{\vec{k}-\vec{p}})+
    f(-\beta\epsilon_{\vec{p}}) f(-\beta\epsilon_{\vec{k}-\vec{p}})]
                                                            \nonumber\\
&& D(\omega - 2\vec k \vec v -mv^2,\vec{k}-2m\vec{v})=\pi\int d^3p\;
   \delta(\omega+\epsilon_{\vec{p}}+\epsilon_{\vec{k}-\vec{p}})\;
   [f(\beta\epsilon_{\vec{p}}) f(\beta\epsilon_{\vec{k}-\vec{p}})-
    f(-\beta\epsilon_{\vec{p}}) f(-\beta\epsilon_{\vec{k}-\vec{p}})]
                                                            \nonumber
\end{eqnarray}
A term like $R^{\star}R\;K(\omega,\vec{k})$ with $K(\omega,\vec{k})>0$ in
the imaginary part of $S_F$ is usually resposible for decoherence. It
suppresses the norm squared of the difference $R$. In our case we have
just a weak inequality $K(\omega,\vec{k})\geq 0$. From its definition
$K>0$ for given $\omega,\vec{k}$ iff there exists such a $\vec{p}$ that
$\omega+\epsilon_{\vec{p}}+\epsilon_{\vec{k}-\vec{p}}=0$. This happens for
$\omega \leq 2\mu-k^2/4m$. In the long wave-length limit, for
$k^2/m\;\ll\mu$ and $\omega\ll\mu$, this condition is satisfied. The long
wave-length modes are subject to decoherence and can be considered as
classical.

  From now on we restrict our attention to the leading order terms
in gradient expansion for $\vec{v}=0$

\begin{eqnarray}\label{S2local}
S^{(2)}_{eff} \approx 
\int d^4x\;
& \{ &
[-\frac{(2\pi)^3}{V}+G(0,\vec{0})] \; ( R^{\star}\Delta \;+\;c.c.\;) +
[\frac{\partial{G}}{\partial\omega}(0,\vec{0})] \;
(-iR^{\star}\partial_t\Delta \;+\;c.c.\;) + \nonumber\\
&& +\sum_{\mu,\nu=1..3}
[\frac{1}{2}\frac{\partial^2 G}
                 {\partial k^{\mu} \partial k^{\nu}}(0,\vec{0}) ]
(R^{\star}\frac{\partial^2\Delta}
               {\partial x^{\mu} \partial x^{\nu}} \;+\;c.c.\;) +
[\frac{\partial D}{\partial\omega}(0,\vec{0})]
(R^{\star} \partial_t \Delta  \;-\;c.c\;)+
i[K(0,\vec{0})]R^{\star}R
\}
\end{eqnarray}
As we can see by inverse Fourier transform the propagator of the density
matrix contains a factor

\begin{equation}\label{decoherence}
e^{ - K(0,\vec{0}) \int dtd^3x\; R^{\star}R }
\end{equation}
which is responsible for decoherence in the long wave-length limit. The
exponent is minus norm squared of the difference $R=\Delta_+-\Delta_-$.
This term suppresses the off-diagonal elements of the density matrix. The
factor (\ref{decoherence}) can be rewritten as

\begin{equation}\label{noise}
\int D\xi^{\star}D\xi
e^{-\int dtd^3x\; 
\left[ \frac{\xi^{\star}\xi}{K(0,\vec{0})} 
       -i\xi^{\star}R -i\xi R^{\star} \right]} \;.
\end{equation}
The noise has the correlation

\begin{equation}
\langle \; \xi^{\star}(t,\vec{x}) \; \xi(t',\vec{x}') \; \rangle = 
K(0,\vec{0}) \; 
\delta(t-t') \;
\delta(\vec{x}-\vec{x}') \;\;.
\end{equation}
Variation of $S_{eff}^{(2)}$ with respect to $R^{\star}$ leads to 

\begin{equation}\label{TDGL}
\mbox{\fboxsep=.1in \framebox{$
\left[ i \frac{\partial G}{\partial\omega}\;-\; 
         \frac{\partial D}{\partial\omega}          \right]\;
\partial_t\Delta\;=\;
\left[ G - \frac{(2\pi)^3}{V} \right] \; \Delta
\;\;+\;\;
\left[
\frac{1}{6}
\frac{\partial^2 G}
     {\partial\vec{k}^2}
\right]\; 
\nabla^2\Delta \;\;+\;\; \xi \;\;
$}}
\end{equation}
where we suppressed the argument $(0,\vec{0})$.  It can be checked that
$0<K(0,\vec{0})=2\beta^{-1}[-\partial_{\omega} D(0,\vec{0}]= \pi N/4$
so that the fluctuation-dissipation relation is satisfied; $N$ is
density of states at the Fermi level. 

The derivation which leads to Eq. (\ref{TDGL}) is valid above critical
temperature, where the expectation value of the order parameter is zero.
It is based on expansion in powers of $\Delta$ around its vanishing
equilibrium value. Below $T_c$ the coefficient $G-(2\pi)^3/V$ is negative
and the system is unstable against spinodal decomposition.

\section{ BELOW $T_c$ }

  In the previous section we made an expansion around vanishing order
parameter $\Delta=0$. Below $T_c$ the order parameter has a nonzero
expectation value which results in a nonzero energy gap for the
quasiparticles. The gap effects can be difficult to generate in
perturbative expansion around $\Delta=0$. That is why we make a
perturbative expansion around uniform and time-independent order
parameter, $\Delta=\Delta_0+\phi$, and work out the effective action up to
second order in $\phi$. A first order term vanishes, if $\Delta_0$
satisfies the standard BCS gap equation. The decoherence and damping
kernels turn out to be nonanalytic at $(\omega,\vec{k})=(0,\vec{0})$ and
it is not possible to make their gradient expansion. Just the
nondissipative (reversible) part of the effective equation can be expanded
and truncated as a partial differential equation; irreversible terms are
nonlocal. The limit of $T=0$ is considered in more detail in the next
section.

  The free electronic Lagrangian for a uniform $\Delta_0$ background is

\begin{equation}
L_0[\psi]=\int d^3x\; 
\left[
\sum_{\alpha=\uparrow,\downarrow} 
\left(
i\psi_{\alpha}^{\dagger}\partial_t\psi_{\alpha}
+\frac{1}{2m}\psi_{\alpha}^{\dagger} \nabla^2 \psi_{\alpha}  
+\mu \psi_{\alpha}^{\dagger}\psi_{\alpha}
\right)+
(2\pi)^3 \Delta_0^{\star} \psi_{\downarrow}\psi_{\uparrow}+
(2\pi)^3 \Delta_0 \psi_{\downarrow}^{\dagger}\psi_{\uparrow}^{\dagger}
\right]
\end{equation}
$\psi$'s can be Fourier transformed as in Eq.(\ref{Fourier}) but $a$'s do
not diagonalize the Hamiltonian. We make the Bogolubov transformation

\begin{eqnarray}
&& a_{\uparrow}(t,\vec{k})=
   u_{\vec{k}}\gamma_{\uparrow}(t,\vec{k})+
   v_{\vec{k}}\gamma_{\downarrow}^{\dagger}(t,-\vec{k})
\;,\nonumber\\
&& a_{\downarrow}(t,\vec{k})=
   u_{\vec{k}}\gamma_{\downarrow}(t,\vec{k})-
   v_{\vec{k}}\gamma_{\uparrow}^{\dagger}(t,-\vec{k})
\;,\nonumber\\
&& |u_{\vec{k}}|^2=\frac{1}{2}\left[
   1+\frac{\epsilon_{\vec{k}}}{e_{\vec{k}}}
   \right]
\;,\nonumber\\
&& |v_{\vec{k}}|^2=\frac{1}{2}\left[
   1-\frac{\epsilon_{\vec{k}}}{e_{\vec{k}}}
   \right]
\;,\nonumber\\
&& 2u_{\vec{k}}v_{\vec{k}}=\frac{(2\pi)^3\Delta_0}{e_{\vec{k}}}
\;,\nonumber\\
&& 
e_{\vec{k}}=\sqrt{(2\pi)^3\Delta_0^{\star}\Delta_0+\epsilon_{\vec{k}}^2}
\;.
\end{eqnarray}
$e_{\vec{k}}$ is the energy of Bogoliubov quasiparticles of momentum
$\vec{k}$. It has a gap $(2\pi)^{3/2}|\Delta_0|$.  After the
transformation the full Lagrangian becomes

\begin{eqnarray}\label{L0}
&&
L[\gamma,\phi]=L_0[\gamma]+L_{\phi}[\phi]+L_{int}[\gamma,\phi]
\;,\nonumber\\
&&
L_0[\gamma]=\int d^3k\;
\left[ \sum_{\alpha=\uparrow,\downarrow}
\left( i\gamma_{\alpha}^{\dagger}\partial_t \gamma_{\alpha}-
       e_{\vec{k}}\gamma_{\alpha}^{\dagger}\gamma_{\alpha} \right)
\right]
\;,\nonumber\\
&&
L_{\phi}[\phi]=\int d^3k\;
\left[
\frac{-(2\pi)^3}{V}
\left(
\Delta_0\phi^{\star}+
\Delta_0^{\star}\phi+
\phi^{\star}\phi
\right)
\right]
\;\nonumber\\
&&
L_{int}[\gamma,\phi]=\int d^3k\;
\left[
\phi^{\star}(t,\vec{k})F_{\vec{k}}[\gamma]+
\phi(t,\vec{k}) F_{\vec{k}}^{\dagger}[\gamma]
\right]
\;.
\end{eqnarray}
We neglected the the ground state energy terms which depend only on
$\Delta_0$. The interaction vertex is

\begin{eqnarray}
F_{\vec{k}}[\gamma]=\int d^3p\;
[&&\;\;
u_{\vec{p}}u_{\vec{k}-\vec{p}}
\gamma_{\downarrow}(t,\vec{p})\gamma_{\uparrow}(t,\vec{k}-\vec{p})-
v_{\vec{p}}v_{\vec{k}-\vec{p}}
\gamma_{\downarrow}^{\dagger}(t,-\vec{p})
\gamma_{\uparrow}^{\dagger}(t,\vec{p}-\vec{k})
\nonumber\\
&&\;\;-v_{\vec{p}}u_{\vec{k}-\vec{p}}
\gamma_{\uparrow}^{\dagger}(t,-\vec{p})
\gamma_{\uparrow}(t,\vec{k}-\vec{p})+
u_{\vec{p}}v_{\vec{k}-\vec{p}}
\gamma_{\downarrow}(t,\vec{p})
\gamma_{\downarrow}^{\dagger}(t,\vec{p}-\vec{k})
\;\;] \;.
\end{eqnarray}
To first order the influence action $S_F$ is

\begin{equation}
S_F^{(1)}=\int dtd^3k\;
\left[
\phi^{\star}_{+}(t,\vec{k})
\langle F_{\vec{k}}[\gamma_+] \rangle -
\phi^{\star}_{-}(t,\vec{k})
\langle F_{\vec{k}}[\gamma_-] \rangle
+ \;\; h.c. \;\; 
\right]\;,
\end{equation}
where $\phi_{\mp}$ lives on (anti-)time ordered branch. This first order
term cancels with the first order term in
$L_{\phi}[\phi_+]-L_{\phi}[\phi_-]$, see Eq.(\ref{L0}), provided that
$\Delta_0$ satisfies the BCS gap equation

\begin{equation}\label{gap}
\frac{2(2\pi)^3}{V}=
\int d^3p\; \frac{\tanh{\frac{\beta e_{\vec{p}}}{2}}}{e_{\vec{p}}} \;.
\end{equation}
There is no linear term in the effective action if the uniform $\Delta_0$
is chosen at thermodynamic equilibrium. The quadratic term of the
effective action is obtained by similar steps as above $T_c$. With the
definition $\phi_{\pm}=\phi\pm R/2$ it becomes

\begin{eqnarray}\label{S2eff<}
S_{eff}^{(2)}=S_{F}^{(2)}+S_{\phi}^{(2)}=
\int d\omega d^3k\; \{\;&&
(\phi R^{\star}+\phi^{\star} R)
\left[ - \frac{(2\pi)^3}{V} + G(\omega,\vec{k}) \right] +
i(\phi^{\star}R-\phi R^{\star}) \; D(\omega,\vec{k}) +
i R^{\star}R \; K(\omega,\vec{k}) +
\nonumber\\&&
\left[ 2\Delta^2_0 \; R^{\star}\phi^{\star} + \;h.c. \right]\;
U(\omega,\vec{k})+
\left[ i(\Delta_0^{\star})^2 \;
        ( \phi^{\star}\phi^{\star}+
          \frac{1}{2}R^{\star}R^{\star} ) \; + \;h.c.\; \right]
\; H(\omega,\vec{k}) \;\} \;\;.
\end{eqnarray}
The kernels $G,D,K,U,H$ are listed in Appendix. $G$ and $U$ can be
expanded in powers of $\omega,\vec{k}$. $D,K,H$ are nonanalytic at
$(\omega,\vec{k})=(0,\vec{0})$, they give rise to nonlocal kernels. 
Any kernel $X$ can be Fourier transformed back

\begin{equation}
X[t-t',\vec{x}-\vec{x}']=
\int \frac{d\omega d^3k}{(2\pi)^4}\;
e^{i\omega(t-t')+i\vec{k}(\vec{x}-\vec{x}')}\;
X[\omega,\vec{k}]\;\;.
\end{equation}
The $R$-quadratic terms in $e^{iS_{eff}^{(2)}}$ can be traded for a
complex noise $\xi$ and real noises $\eta_{\star},\eta$,

\begin{eqnarray}
&&e^{-\int d^4xd^4x'\;
      \left[
      R^{\star}(x)K(x-x')R(x')+
      \frac{1}{2}(\Delta_0^{\star})^2 R^{\star}(x)H(x-x')R^{\star}(x')+
      \frac{1}{2}(\Delta_0)^2R(x)H(x-x')R(x')
      \right] } =
\nonumber\\
&&\int D\xi^{\star} D\xi D\eta_{\star} D\eta \;\;
e^{-\int d^4xd^4x'\; 
   \left[    
   \xi^{\star} K^{-1} \xi+
   \eta_{\star} \left(\frac {H^{-1}} 2\right) \eta_{\star}+
   \eta \left(\frac {H^{-1}} 2\right) \eta   
   \right]}
e^{i \int d^4x\;
     \left[
     (\xi R^{\star} +\;c.c.\;) +
     i \eta_{\star} R^{\star} \Delta_0^{\star}+
     \eta R \Delta_0 
   \right]}   \;.
\end{eqnarray}
Grandient expansion of $G,U$ and subsequent variation of the effective
action with respect to $R^{\star}$ gives an effective equation

\begin{eqnarray}
&&\int d^4x' \; iD(x-x') \; \phi(x') -
\left[ \frac{\partial G}{\partial \omega} \right]\; 
i\partial_t\phi+
\left[ \frac{1}{6}\frac{\partial^2 G}{\partial\vec{k}^2} \right]\;
\nabla^2\phi+
\left[ \frac{1}{2}\frac{\partial^2 G}{\partial \omega^2} \right]\;
\partial_t^2\phi+
\left[ \frac{\partial^2 U}{\partial \omega^2} \right]\; 
\Delta_0^2 \partial_t^2 \phi^{\star}+
\left[ \frac{1}{3}\frac{\partial^2 U}{\partial\vec{k}^2} \right]\;
\Delta_0^2 \nabla^2 \phi^{\star} =  \nonumber\\
&&=\left[ G-\frac{(2\pi)^3}{V} \right]\; \phi+
\left[ 2U \right] \Delta_0^2 \; \phi^{\star}+
\xi+
i\Delta_0^{\star} \; \eta_{\star}
\end{eqnarray}
where we kept derivatives up to second order. The kernels and their
derivatives in square brackets are evaluated at $(\omega,\vec{k})=(0,\vec
0)$. In the special case of $T=0$, which is considered in the next
section, one can reconstruct from this linearized equation a
nonperturbative equation for $\Delta(t,\vec{x})$.

\section{ SMALL TEMPERATURE OR $T\ll |\Delta_0|$ }

  The effective equation takes a particularly simple form in the low
frequency and long wavelength limit close to zero temperature.  The
nonanalytic kernels (\ref{K0},\ref{D0},\ref{H0}) can be ignored provided
that temperature is small as compared to the quasiparticle gap and
$|\omega|$ is less than twice the gap. In this regime there is no
decoherence or noise from pair-breaking fluctuations. Nontrivial
contributions come from the analytic kernels (\ref{G0},\ref{U0}). Both of
them can be gradient expanded. When we restrict ourselves to leading time
derivative and potential term, the effective equation for small
fluctuations will take the form

\begin{equation}\label{eqphi}
  \left[ -\frac{\partial G}{\partial\omega}(0,\vec 0)
  \right]
  i\partial_t\phi 
= 
\frac{(2\pi)^3\;\phi}{V} +
\phi\int \frac{ d^3p }
              { 2\; e_{\vec{p}}[\Delta_0^{\star}\Delta_0] } +
\Delta_0\phi\frac{\partial}{\partial\Delta_0}
 \int \frac{ d^3p }{ 2\; e_{\vec{p}}[\Delta_0^{\star}\Delta_0] } +
\Delta_{0}\phi^{\star}\frac{\partial}{\partial\Delta_0^{\star}}
 \int \frac{ d^3p }{ 2\; e_{\vec{p}}[\Delta_0^{\star}\Delta_0]
}\;,
\end{equation} 
We remind explicitly that $e$'s depend on $\Delta_0$. This equation
follows from the effective Lagrangian

\begin{equation}\label{tildetildeLeff}
\tilde{\tilde{L}}_{eff}=\int d^3x\;
\left\{
F[\rho]\;
\left( \frac{i\partial_t\Delta}{\Delta}+c.c. \right) 
-V_{eff}[\rho]
\right\} \;.
\end{equation}
The real coefficients are
functions of $\rho\stackrel{def}{=}\Delta^{\star}\Delta$:

\begin{eqnarray}
&& F[\rho]=
\int\frac{d^3p\;\epsilon_{\vec{p}}}
         {4(2\pi)^3}\;
    \frac{1}{e_{\vec{p}}(t,\vec{x})}
\;,\nonumber\\
&& V_{eff}[\rho]=
\frac{(2\pi)^3 \; \rho(t,\vec{x})}{V}-
\int\frac{d^3p}{(2\pi)^3}\;
e_{\vec{p}}(t,\vec{x})
\;.\nonumber\\
\end{eqnarray}
The coefficients depend on $\rho$ through the energies $e$ which appear in
the definitions (\ref{G0},\ref{U0}) but with the energies generalized to

\begin{equation}
e_{\vec{p}}(t,\vec{x})=\sqrt{ (2\pi)^3
\rho(t,\vec{x}) + \epsilon_{\vec{p}}^2 } \;.
\end{equation}
The equation (\ref{eqphi}) follows from the Lagrangian
(\ref{tildetildeLeff}) as

\begin{equation}
0 = \left( \phi \; \frac{\delta}{\delta\Delta} +
           \phi^{\star} \; \frac{\delta}{\delta\Delta^{\star}} \right)\;
    \frac{\delta L_{eff}}{\delta\Delta^{\star}} \;\;,
\end{equation}
where the second variation is around the constant $\Delta_0$ background.
The leading phase gradient terms can be reconstructed using Galilean
symmetry

\begin{equation}\label{tildeLeff}
\tilde L_{eff}=\int d^3x\;
\left\{
-2F[\rho]\;
\left(\; \partial_t \chi \;+\;
         \frac{(\nabla\chi)^2}{4m}                \right)
-V_{eff}[\rho]
\right\}   \;\;,
\end{equation}
where $\Delta\stackrel{def}{=}\rho^{1/2}\exp i\chi $. This however still
leaves undetermined the Galilean invariant term $ \sim \int
d^3x\;(\nabla\rho)^2 $. According to \cite{kp,ao} the perturbative version
of such a term is $-F(\rho_0)[\nabla(\delta\rho^{1/2})]^2/12m\rho_0$,
where $\rho_0=|\Delta_0|^2$. With this term included the effective
Lagrangian reads

\begin{equation}\label{Leff}
\mbox{\fboxsep=.1in \framebox{$
L_{eff}=\int d^3x\;
\left\{
-2F[\rho]\;
\left(\; \partial_t \chi \;+\;                
         \frac{(\nabla\chi)^2}{4m} \;+\;
         \frac{(\nabla \rho^{1/2})^2}{12m\rho}               
\right)
-V_{eff}[\rho]
\right\}   \;\;,
$}}
\end{equation}
plus higher order derivative terms.

  The Euler-Lagrange which follows from (\ref{Leff}) for a constant
$\Delta\;$, $\delta V_{eff}/\delta\Delta^{\star}=0$, is identical to the
zero temperature version of the gap equation (\ref{gap}).

  We observe that $\;4F[\rho]=\partial V_{eff}[\rho]/\partial\mu$ and as
such is identified as density of particles. This observation is consistent
with Ref. \cite{stone2}. In contrast to \cite{stone2} our Lagrangian
(\ref{tildeLeff}) is valid even for nonuniform $| \Delta |$.

  The effective Lagrangian contains a first order time derivative term
$\sim\partial_t\chi$. It is not a total time derivative. This term was
neglected in some early treatments because its coefficient $F[\rho]$
vanishes with the usual approximation $\int d^3p \approx 4\pi N
\int^{+\omega_D}_{-\omega_D} d\epsilon_p$, where $N$ is density of states
at the Fermi level and $\omega_D$ is the Debye frequency.  In
\cite{stone2,ao} the term is reconstructed with Galilean invariance but
only as a total time derivative. This term implies that at zero
temperature in clean superconductor a moving vortex is feeling the Magnus
force transversal to its velocity.

  The gradient terms can not be rearranged as $|\nabla\Delta|^2$, there is
different stiffness for phase and modulus fluctuations. This is in
contrast to the situation above $T_c$, compare Eq.(\ref{TDGL}), and to the
Gross-Pitaievskii equation.

  The effective Lagrangian is consistent with the previous work on this
topic up to discrepancies which have been identified and explained. The
new ficture is the $\rho$ dependence in $F(\rho)$ and the nonpolynomial
character of $V_{eff}(\rho)$.

\section*{ CONCLUDING REMARKS }

 We made a thourough survey of decoherence and noise in neutral s-wave
superconductor. The dynamics above critical temperature is of Langevin
type. In the long wavelength limit the dissipation kernel can be
approximated by a simple friction term and the source of fluctuations by
white gaussian noise. Long wavelength modes decohere and are effectively
classical. This description survives below $T_c$ as long as the gap
remains small as compared to temperature. As the gap developes with
decreasing temperature the dissipation kernel and noise correlations
become essentially nonlocal - they can not be expanded in powers of
derivatives. The decoherence also has a nonlocal kernel. As the
temperature becomes small as compared to the gap noise, dissipation and
decoherence become exponentially supressed except for frequencies of at
least twice the gap. For low temperatures we obtain a quantum system.

 The quasiparticle gap is the key factor to suppress decoherence and
noise. This fact can be easily overlooked if the effective theory is
constructed by expansion around $\Delta=0$ in all temperature regimes. It
is difficult to generate gap effects in the leading orders of such
expansion.

 At near zero temperature we get an effective quantum Lagrangian in terms
of the order parameter $\Delta$. The Lagrangian contains a first order
time derivative term. This term is responsible for the Magnus force acting
on a vortex moving with respect to the superfluid and phenomena such as
parallel motion of vortex-antivortex pair or vortex pair rotation around
its center of mass. The existence of this term is also an essential
assumption for the phenomenological theory of Josephson junctions.


\acknowledgements I would like to thank Wojtek Zurek for discussions and
Diego Dalvit for his critical interest in this work. Work supported by KBN
grant 2 P03B 008 15.



\section*{Appendix}

   The nonlocal kernels encountered below $T_c$ can be divided into those
which have a well defined gradient expansion and those which are
nonanalytic at $(\omega,\vec{k})=(0,\vec{0})$. The {\bf analytic
kernels} are $G$ and $U$.

\begin{eqnarray}\label{G}
G(\omega,\vec{k})=\;P.V.\;\int d^3p\;\{&&
\left[ f(\beta e_{\vec{p}})f(\beta e_{\vec{k}-\vec{p}}) -
       f(-\beta e_{\vec{p}})f(-\beta e_{\vec{k}-\vec{p}}) \right]
\left[ \frac{-|u_{\vec{p}}|^2|u_{\vec{k}-\vec{p}}|^2}
            {\omega+e_{\vec{p}}+e_{\vec{k}-\vec{p}}} +
       \frac{|v_{\vec{p}}|^2|v_{\vec{k}-\vec{p}}|^2}
            {\omega-e_{\vec{p}}-e_{\vec{k}-\vec{p}}}      \right]+
\nonumber\\
&&
\left[ f(\beta e_{\vec{p}})f(-\beta e_{\vec{k}-\vec{p}}) -
       f(-\beta e_{\vec{p}})f(\beta e_{\vec{k}-\vec{p}}) \right]
\left[ \frac{|v_{\vec{p}}|^2|u_{\vec{k}-\vec{p}}|^2}  
            {\omega-e_{\vec{p}}+e_{\vec{k}-\vec{p}}} +
       \frac{|u_{\vec{p}}|^2|v_{\vec{k}-\vec{p}}|^2}
            {\omega+e_{\vec{p}}-e_{\vec{k}-\vec{p}}} \right] \} 
\end{eqnarray}
As a selfconsistency check we note that for $\Delta_0=0$ the $G$ kernel
coincides with the $G$ in Eq. (\ref{S2eff}).

\begin{eqnarray}\label{U}
U(\omega,\vec{k})=\frac{(2\pi)^3}{4}\;P.V.\;\int d^3p\;\{&&
\left[ f(\beta e_{\vec{p}})f(\beta e_{\vec{k}-\vec{p}}) -
       f(-\beta e_{\vec{p}})f(-\beta e_{\vec{k}-\vec{p}}) \right]
\left[ \frac{e_{\vec{k}-\vec{p}}+e_{\vec{p}}}
            {e_{\vec{p}}e_{\vec{k}-\vec{p}}
             [(e_{\vec{p}}+e_{\vec{k}-\vec{p}})^2-\omega^2]} \right]+
\nonumber\\
&&\left[ f(\beta e_{\vec{p}})f(-\beta e_{\vec{k}-\vec{p}}) -
         f(-\beta e_{\vec{p}})f(\beta e_{\vec{k}-\vec{p}}) \right]
  \left[ \frac{e_{\vec{k}-\vec{p}}-e_{\vec{p}}}
              {e_{\vec{p}}e_{\vec{k}-\vec{p}}
               [(e_{\vec{p}}+e_{\vec{k}-\vec{p}})^2-\omega^2]} \right]\}
\end{eqnarray}
At {\bf zero temperature } these kernels simplify to

\begin{eqnarray}
&& G(\omega,\vec{k}) \stackrel{\beta=\infty}{=}\;P.V.\;\int d^3p\;
   \left[ \frac{|u_{\vec{p}}|^2|u_{\vec{k}-\vec{p}}|^2}
               {\omega+e_{\vec{p}}+e_{\vec{k}-\vec{p}}} +
          \frac{-|v_{\vec{p}}|^2|v_{\vec{k}-\vec{p}}|^2}
               {\omega-e_{\vec{p}}-e_{\vec{k}-\vec{p}}}      \right]
\label{G0}\\
&& U(\omega,\vec{k}) \stackrel{\beta=\infty}{=}\;P.V.\; 
   \frac{(2\pi)^3}{4}\int d^3p\;
   \left[ \frac{e_{\vec{k}-\vec{p}}+e_{\vec{p}}}
               {e_{\vec{p}}e_{\vec{k}-\vec{p}}  
               [\omega^2-(e_{\vec{p}}+e_{\vec{k}-\vec{p}})^2]}\right]
\label{U0}
\end{eqnarray}
The {\bf nonanalytic kernels } are as follows

\begin{eqnarray}\label{K}
K(\omega,\vec{k})=\pi\int d^3p\; \{ 
&& \left[ f(\beta e_{\vec{p}})f(\beta e_{\vec{k}-\vec{p}}) +
          f(-\beta e_{\vec{p}})f(-\beta e_{\vec{k}-\vec{p}}) \right]\times
\nonumber\\
&& \left[ |u_{\vec{p}}|^2|u_{\vec{k}-\vec{p}}|^2 \;
          \delta(\omega+e_{\vec{p}}+e_{\vec{k}-\vec{p}}) +
          |v_{\vec{p}}|^2|v_{\vec{k}-\vec{p}}|^2 \;
          \delta(\omega-e_{\vec{p}}-e_{\vec{k}-\vec{p}}) \right] +
\nonumber\\
&& \left[ f(\beta e_{\vec{p}})f(-\beta e_{\vec{k}-\vec{p}}) +
          f(-\beta e_{\vec{p}})f(\beta e_{\vec{k}-\vec{p}}) \right]\times
\nonumber\\
&& \left[ |v_{\vec{p}}|^2|u_{\vec{k}-\vec{p}}|^2 \;
          \delta(\omega-e_{\vec{p}}+e_{\vec{k}-\vec{p}}) +
          |u_{\vec{p}}|^2|v_{\vec{k}-\vec{p}}|^2 \;
          \delta(\omega+e_{\vec{p}}-e_{\vec{k}-\vec{p}}) \right] \}       
\end{eqnarray}
\begin{eqnarray}\label{D}
D(\omega,\vec{k})=\pi\int d^3p\; \{
&& \left[ f(\beta e_{\vec{p}})f(\beta e_{\vec{k}-\vec{p}}) -
          f(-\beta e_{\vec{p}})f(-\beta e_{\vec{k}-\vec{p}}) \right]\times
\nonumber\\
&& \left[ |u_{\vec{p}}|^2|u_{\vec{k}-\vec{p}}|^2 \;
          \delta(\omega+e_{\vec{p}}+e_{\vec{k}-\vec{p}}) -
          |v_{\vec{p}}|^2|v_{\vec{k}-\vec{p}}|^2 \;
          \delta(\omega-e_{\vec{p}}-e_{\vec{k}-\vec{p}}) \right] +
\nonumber\\   
&& \left[ f(\beta e_{\vec{p}})f(-\beta e_{\vec{k}-\vec{p}}) -
          f(-\beta e_{\vec{p}})f(\beta e_{\vec{k}-\vec{p}}) \right]\times
\nonumber\\
&& \left[ |v_{\vec{p}}|^2|u_{\vec{k}-\vec{p}}|^2 \;
          \delta(\omega-e_{\vec{p}}+e_{\vec{k}-\vec{p}}) -
          |u_{\vec{p}}|^2|v_{\vec{k}-\vec{p}}|^2 \;   
          \delta(\omega+e_{\vec{p}}-e_{\vec{k}-\vec{p}}) \right] \}
\end{eqnarray}
Again we note that for $\Delta_0=0$ these $K$ and $D$ kernels match the
corresponding ones in Eq.(\ref{S2eff}).

\begin{eqnarray}\label{H}
H(\omega,\vec{k})=\frac{\pi(2\pi)^3}{8}\int d^3p\;
\frac{1}{e_{\vec{p}}e_{\vec{k}-\vec{p}}}\;\{
&& \left[ f(\beta e_{\vec{p}})f(\beta e_{\vec{k}-\vec{p}}) +
          f(-\beta e_{\vec{p}})f(-\beta e_{\vec{k}-\vec{p}}) \right]\times
\nonumber\\
&& \left[ \delta(\omega+e_{\vec{p}}+e_{\vec{k}-\vec{p}}) +
          \delta(\omega-e_{\vec{p}}-e_{\vec{k}-\vec{p}}) \right] +
\nonumber\\
&& \left[ f(\beta e_{\vec{p}})f(-\beta e_{\vec{k}-\vec{p}}) +
          f(-\beta e_{\vec{p}})f(\beta e_{\vec{k}-\vec{p}}) \right]\times
\nonumber\\
&& \left[ \delta(\omega-e_{\vec{p}}+e_{\vec{k}-\vec{p}}) +
          \delta(\omega+e_{\vec{p}}-e_{\vec{k}-\vec{p}}) \right] \}
\end{eqnarray}
At {\bf zero temperature } the nonanalytic kernels become

\begin{eqnarray}
&&K(\omega,\vec{k})\stackrel{\beta=\infty}{=}\pi\int d^3p\; 
  \left[ |u_{\vec{p}}|^2|u_{\vec{k}-\vec{p}}|^2 \;
         \delta(\omega+e_{\vec{p}}+e_{\vec{k}-\vec{p}}) +
         |v_{\vec{p}}|^2|v_{\vec{k}-\vec{p}}|^2 \;
         \delta(\omega-e_{\vec{p}}-e_{\vec{k}-\vec{p}}) \right] 
\label{K0}\\
&&D(\omega,\vec{k})\stackrel{\beta=\infty}{=}-\pi\int d^3p\; 
  \left[ |u_{\vec{p}}|^2|u_{\vec{k}-\vec{p}}|^2 \;
         \delta(\omega+e_{\vec{p}}+e_{\vec{k}-\vec{p}}) -
         |v_{\vec{p}}|^2|v_{\vec{k}-\vec{p}}|^2 \;
         \delta(\omega-e_{\vec{p}}-e_{\vec{k}-\vec{p}}) \right] 
\label{D0}\\
&& H(\omega,\vec{k})\stackrel{\beta=\infty}{=}
   \frac{\pi(2\pi)^3}{8}\int d^3p\;
   \frac{1}{e_{\vec{p}}e_{\vec{k}-\vec{p}}}\;
   \left[ \delta(\omega+e_{\vec{p}}+e_{\vec{k}-\vec{p}}) +
          \delta(\omega-e_{\vec{p}}-e_{\vec{k}-\vec{p}}) \right]
\label{H0}
\end{eqnarray}

\end{document}